# Skin Lesion Segmentation and Classification for ISIC 2018 Using Traditional Classifiers with Hand-Crafted Features

Russell C. Hardie, Redha Ali, Manawaduge Supun De Silva, and Temesguen Messay Kebede

Signal and Image Processing Lab
Department of Electrical and Computer Engineering
University of Dayton, 300 College Park, Dayton OH 45469-0232
Email: rhardie1@udayton.edu

7/18/2018

## Abstract

This paper provides the required description of the methods used to obtain submitted results for Task1 and Task 3 of ISIC 2018: Skin Lesion Analysis Towards Melanoma Detection [1]. The results have been created by a team of researchers at the University of Dayton Signal and Image Processing Lab. In this submission, traditional classifiers with hand-crafted features are utilized for Task 1 and Task 3. Our team is providing additional separate submissions using deep learning methods for comparison.

## 1. Introduction

The International Skin Imaging Collaboration (ISIC) is a recurring challenge to develop image analysis tools for the automated segmentation and diagnosis of skin lesions with the aim of accurate melanoma detection from dermascopic images [1]. The challenge is broken into three tasks: 1. Lesion Segmentation, 2. Lesion Attribute Detection, and 3. Disease Classification [1]. This paper describes methods used for a Task 1 and Task 3 submission for ISIC 2018. The approach taken in this paper is to use traditional classifier methods with hand-crafted features. We believe this provides an interesting comparison with deep learning methods. The current authors are also preparing results with deep learning methods for a separate submission.

## 2. Task 1 Methodology

Task 1 involves the segmentation of lesions from lesion images acquired with a variety of dermatoscope types. The method used in this submission processes the images in RGB space. The approach begins with the design of a color classifier to distinguish lesion tissue from normal skin tissue based exclusively on RGB color vectors. Gaussian mixture models (GMMs) are used to model the probability density functions of the tissue types. The training data provided is used to estimate the GMM probability density functions for the tissue types. A Bayesian classifier framework is used to estimate the posterior probability of being a lesion pixel for all pixels in a given image. The segmentation threshold for each image is adaptively selected using a support vector machine (SVM) regression algorithm [2]. The SVM regression is trained to predict the

Jaccard index (i.e., overlap score) with respect to the truth in a fashion similar to that proposed by two of the current authors in [2]. The threshold that maximizes the regression network predicted overlap score is selected to provide the segmentation output. Morphological operations are used after the thresholding to fine-tune the segmentation mask. These morphological operations include opening, closing and filling holes.

## 3. Task 3 Methodology

The classification method for lesion diagnosis in this submission is accomplished using a SVM classifier with 200 hand-crafted features. The features are computed from the RGB image along with the lesion segmentation mask obtained using the method in Task 1 above. The features employed are similar to those used in [3-5]. However, here they are computed for each of the three color channels and concatenated and then fed into the SVM classifier.

## 4. Experimental Results

### 4.1 Task 1

Preliminary results for Task 1 have been obtained using the provided training imagery. Training on the odd numbered cases and testing on the even numbered cases we obtain the overlap scores shown in Fig. 1. The mean overlap score on the testing data is 0.776, but drops to 0.701 after zeroing those less than 0.65, as required by official scoring. In this experiment, 17.66% of the testing cases fall below the 0.65 threshold. The current algorithm appears to have the most difficulty with the large spread-out lesions that nearly fill the image. This is understandable given the limited amount of spatial information utilized by the simple color-based pixel segmentation algorithm. On the validation set for ISIC 2018, this method received a score of 0.663 for Task 1.

Some example segmentations are shown in Figs. 2-5. In these images the red contour is the truth mask and the green contour is the predicted lesion boundary using the proposed method. Figure 2 shows two examples with a high overlap score (i.e., $> 0.90$). Figure 3 shows examples where the truth masks appears to very conservative and include a significant amount of what appears to be normal tissue. This hurts the overlap score of the predicted segmentation, even though the algorithm results appear to be quite good. Figure 4 shows two failed cases (i.e., overlap $< 0.65$). Here the predicted segmentations are too small. Finally, Fig. 5 shows two examples where the proposed segmentation algorithm does a fairly good job dealing with a significant amount of hair on the skin.

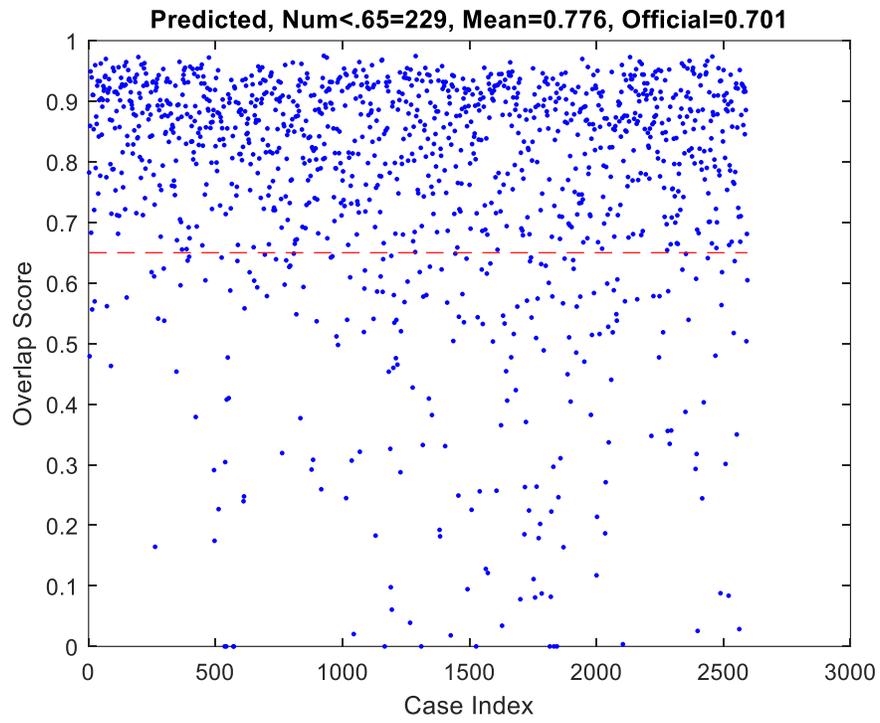

**Fig. 1:** Task 1 performance analysis training on the odd cases and testing on the even cases.

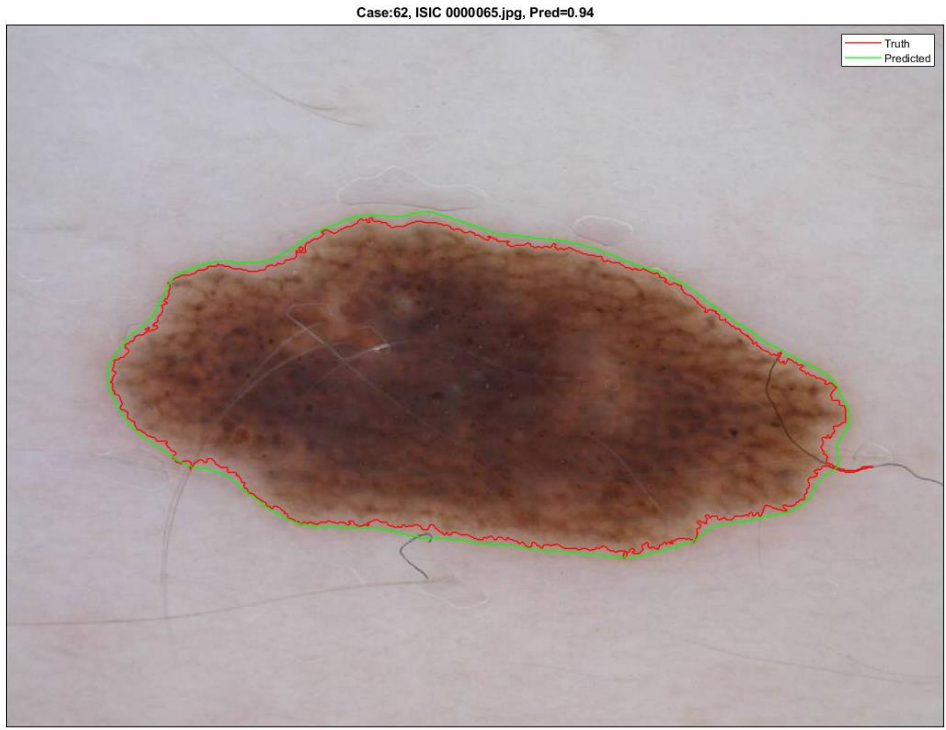

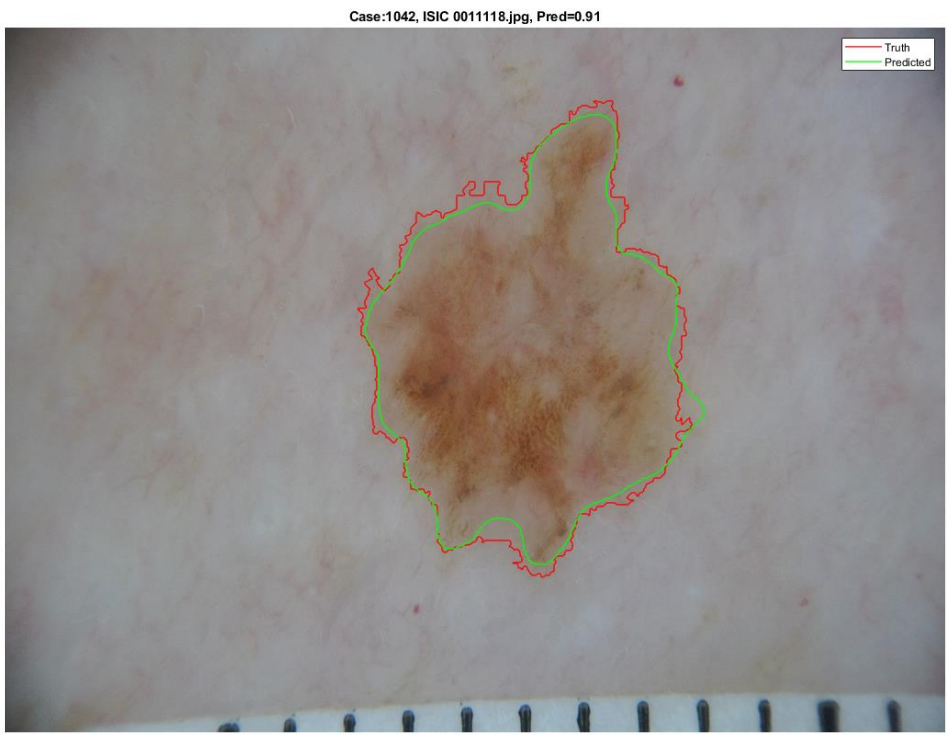

**Fig 2:** Two example segmentations with > 0.90 overlap score.

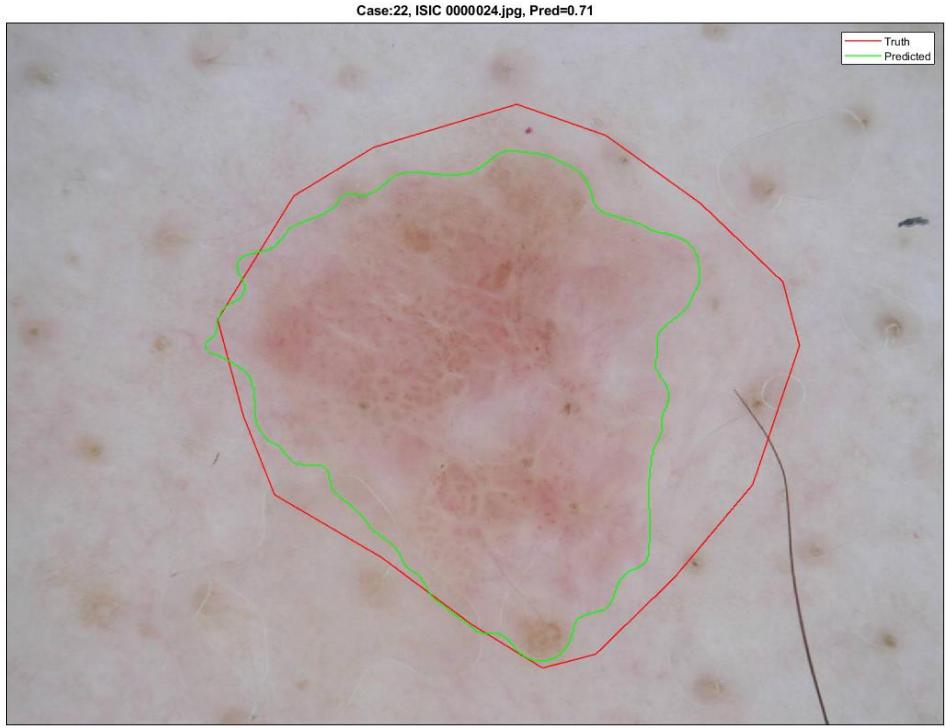

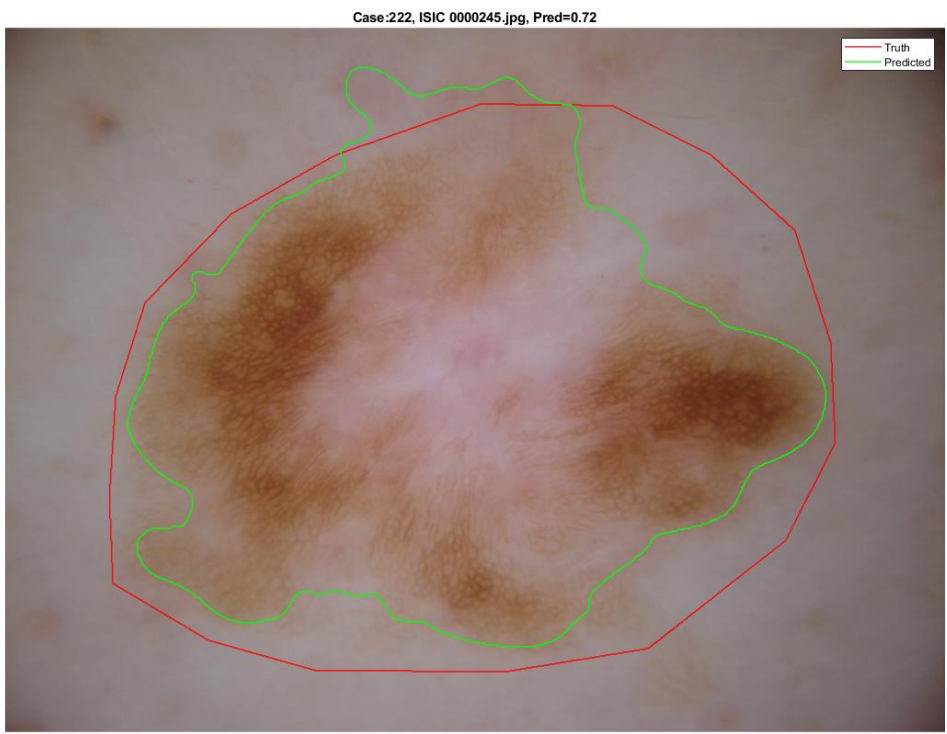

**Fig. 3:** Two examples where the truth mask appears to include a significant amount of normal tissue, causing a reduced score.

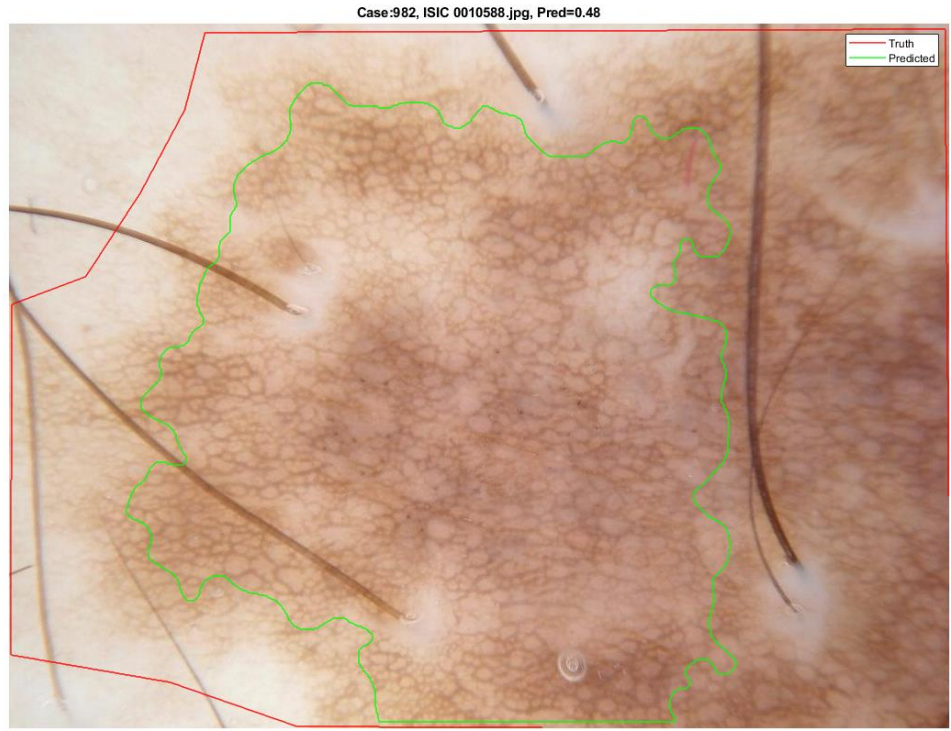
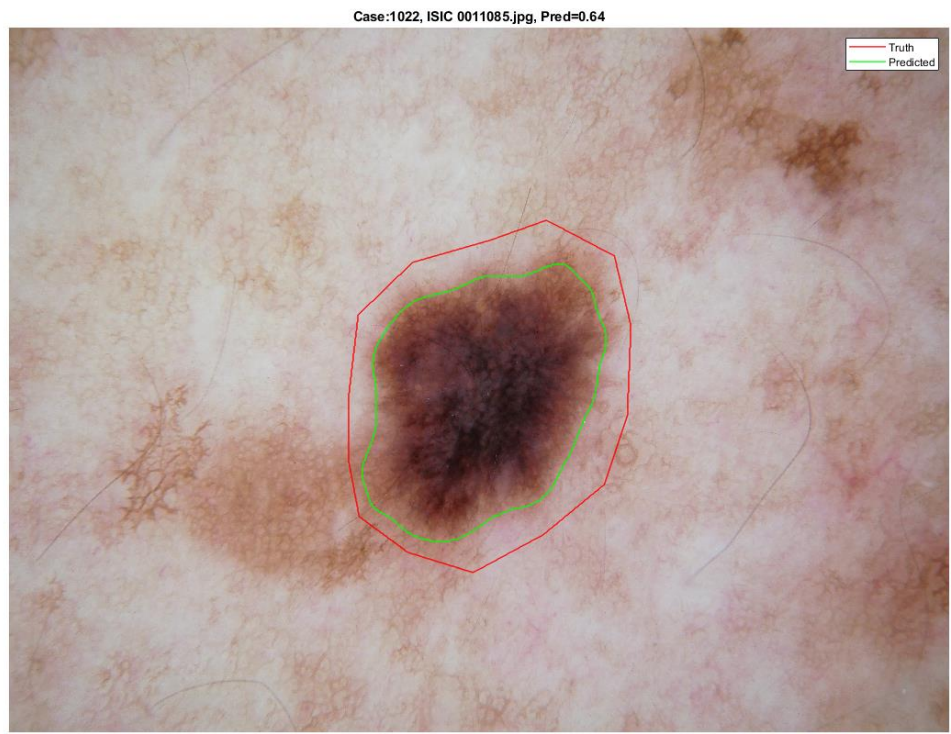

**Fig. 4:** Two examples of failed cases with overlap score < 0.65.

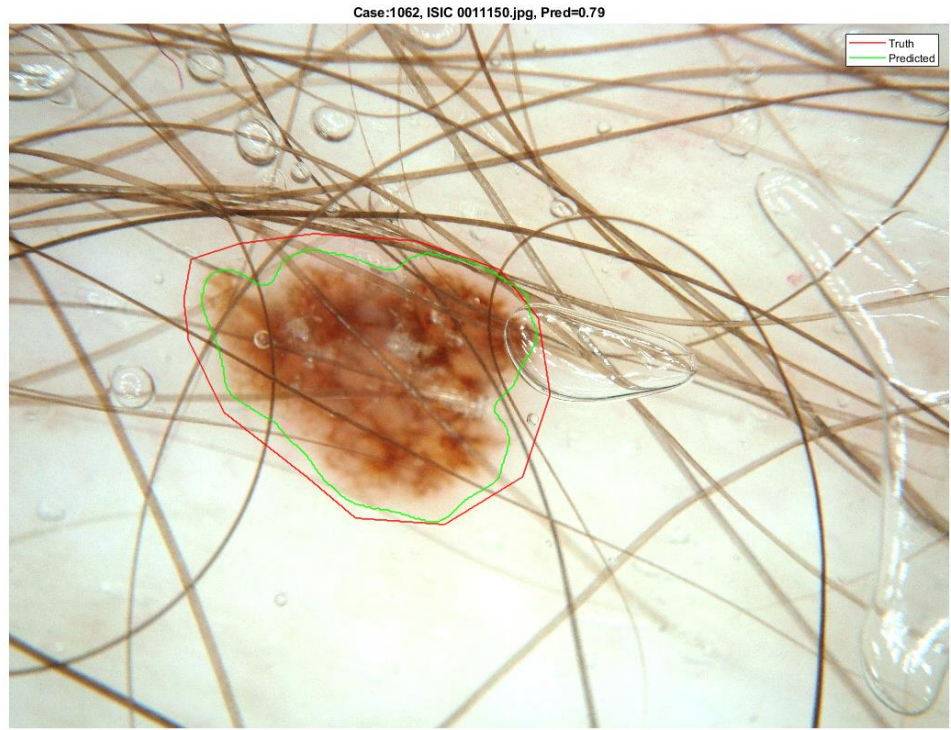

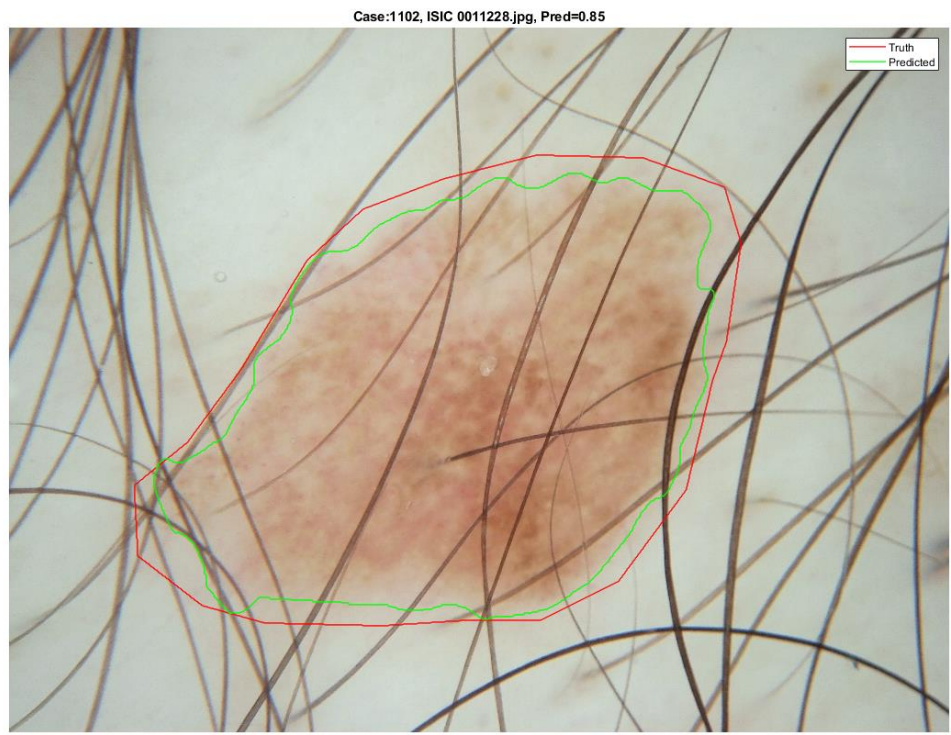

**Fig. 5:** Two examples of the algorithm successfully handing a significant amount of hair.

## 4.2 Task 3

Performance analysis for Task 3 is evaluated using 5-fold cross-validation on the provided training imagery. The 5-fold cross-validation classifier confusion matrix for the 7 classes is shown in Table 1. Performance metrics derived from the confusion matrix are shown in Table 2. The class-averaged recall (same as class-averaged sensitivity) is 0.7303. For the ISIC 2018 validation data, this method received a score of 0.772.

**Table 1:** Confusion matrix for Task 3 using 5-fold cross-validation on the training data.

|  | MEL | NV | BCC | AKIEC | BKL | DF | VASC |
|---|---|---|---|---|---|---|---|
| **MEL** | 783 | 122 | 32 | 44 | 117 | 13 | 2 |
| **NV** | 788 | 5233 | 184 | 85 | 354 | 47 | 14 |
| **BCC** | 11 | 11 | 414 | 39 | 35 | 3 | 1 |
| **AKIEC** | 15 | 10 | 44 | 229 | 27 | 1 | 1 |
| **BKL** | 142 | 82 | 70 | 70 | 717 | 17 | 1 |
| **DF** | 2 | 7 | 9 | 13 | 4 | 80 | 0 |
| **VAS** | 6 | 13 | 8 | 1 | 2 | 2 | 110 |

**Table 2:** Performance metrics derived from the confusion matrix in Table 1.

|  | MEL | NV | BCC | AKIEC | BKL | DF | VASC |
|---|---|---|---|---|---|---|---|
| **Accuracy** | 0.8708 | 0.8286 | 0.9554 | 0.9651 | 0.9080 | 0.9882 | 0.9949 |
| **Error Rate** | 0.1292 | 0.1714 | 0.0446 | 0.0349 | 0.0920 | 0.0118 | 0.0051 |
| **Sensitivity** | 0.7035 | 0.7805 | 0.8054 | 0.7003 | 0.6524 | 0.6957 | 0.7746 |
| **Specificity** | 0.8917 | 0.9260 | 0.9635 | 0.9740 | 0.9395 | 0.9916 | 0.9981 |
| **Precision** | 0.4482 | 0.9553 | 0.5440 | 0.4761 | 0.5709 | 0.4908 | 0.8527 |
| **Recall** | 0.7035 | 0.7805 | 0.8054 | 0.7003 | 0.6524 | 0.6957 | 0.7746 |

## 5. Conclusions

We have provided an ISIC 2018 submission for Task 1 and Task 3 [1] using a traditional processing approach. For Task 1 we use an RGB Bayes classifier guided by a regression network to adaptively set the threshold [2]. Our training/testing analysis produces an average overlap score of 0.701 (after zeroing any overlap less than 0.65).

For Task 3, we use hand-crafted features that make use of the lesion segmentations from the Task 1 algorithm. The features are fed into an SVM classifier. The average class recall (or sensitivity) using 5-fold cross-validation on the provided training imagery is 0.7303.